\documentclass{aa}  

\usepackage{graphicx}
\usepackage{txfonts}
\usepackage{natbib,twoopt}
\usepackage[breaklinks=true]{hyperref} 
\bibpunct{(}{)}{;}{a}{}{,}             

\usepackage{ulem}

\usepackage{pifont}
\newcommand{\cmark}{\ding{51}}%
\newcommand{\xmark}{\ding{55}}%

\begin{document}

   \title{Image Quality Assessment for Full-Disk Solar Observations with Generative Adversarial Networks}
   \titlerunning{Solar Image Quality Assessment with GANs}

   \author{
   R. Jarolim\inst{1} \and 
   A. M. Veronig\inst{1,2} \and 
   W. Pötzi\inst{2} \and 
   T. Podladchikova\inst{3}
   }

   \institute{University of Graz, Institute of Physics, Universitätsplatz 5, 8010 Graz, Austria
         \and University of Graz, Kanzelhöhe Observatory for Solar and Environmental Research, Kanzelhöhe 19, 9521 Treffen am Ossiacher See, Austria \and
         Skolkovo Institute of Science and Technology, Bolshoy Boulevard 30, bld. 1, Moscow 121205, Russia\\ \email{robert.jarolim@uni-graz.at}
             }

   \date{Received 18 June 2020; accepted 24 August 2020}

 
  \abstract
    {Within the last decades, solar physics has entered the era of big data and the amount of data being constantly produced from ground- and space-based observatories can no longer be purely analyzed by human observers.}
   {In order to assure a stable series of recorded images of sufficient quality for further scientific analysis, an objective image quality measure is required. Especially when dealing with ground-based observations, which are subject to varying seeing conditions and clouds, the quality assessment has to take multiple effects into account and provide information about the affected regions. The automatic and robust identification of quality-degrading effects is a critical task, in order to maximize the scientific return from the observations and to allow for event detections in real-time. In this study, we develop a deep learning method that is suited to identify anomalies and provide an image quality assessment of solar full-disk H$\alpha$ filtergrams. The approach is based on the structural appearance and the true image distribution of high-quality observations.}
   {We employ a neural network with an encoder-decoder architecture to perform an identity transformation of selected high-quality observations. The encoder network is used to achieve a compressed representation of the input data, which is reconstructed to the original by the decoder. We use adversarial training to recover truncated information based on the high-quality image distribution. When images with reduced quality are transformed, the reconstruction of unknown features (e.g., clouds, contrails, partial occultation) shows deviations from the original. This difference is used to quantify the quality of the observations and to identify the affected regions. In addition, we present an extension of this architecture by using also low-quality samples in the training step, which takes the characteristics of both quality domains into account and improves the sensitivity for minor image quality degradation.}
   {We apply our method to full-disk H$\alpha$ filtergrams from Kanzelh\"ohe Observatory recorded during 2012-2019 and demonstrate its capability to perform a reliable image quality assessment for various atmospheric conditions and instrumental effects. Our quality metric achieves an accuracy of 98.5\% in distinguishing observations with quality-degrading effects from clear observations and provides a continuous quality measure which is in good agreement with the human perception.}
   {The developed method is capable of providing a reliable image quality assessment in real-time, without the requirement of reference observations. Our approach has the potential for further application to similar astrophysical observations and requires only little effort of manual labeling.}


   \maketitle
%

\section{Introduction}
\label{section:introduction}

Modern solar observations are carried out in an autonomous way, covering multiple filter bands and producing a high cadence output stream which needs to be accessible for monitoring and scientific use within minutes \citep{harvey1996global,pesnell2012sdo,potzi2018event}. While the continuous observation provides a clear benefit of a permanent monitoring of the Sun, automatic and robust methods are necessary to ensure the image quality in the large data streams, before they are passed on for further analysis. This is in particular necessary for ground-based observations, which are subject to varying seeing condition, as well as applying event detection (e.g., flares, filament eruptions) on real-time data \citep{potzi2015real, veronig2016spaceweather}.

Ground-based observations provide the advantage of upgradability and cost efficiency as compared to space-based observations, but have to overcome the limiting factors of atmospheric turbulence. Adaptive optics systems \citep{rimelle2011solaradaptive} and post-facto corrections \citep{lofdahl2007restoration} have shown the ability to correct for local seeing conditions. For autonomous full-disk observations these methods can not be applied, since they rely to some extend on human supervision \citep{woger2008speckle, lofdahl2007restoration}. In addition, the daily observation schedule, varying seeing conditions and the possible presence of clouds lead to unavoidable gaps in the continuous observation series. The use of ground-based network telescopes has shown the capability to provide a continuous data stream and can mitigate the impact of local effects, as can be seen e.g. from operating ground-based networks for observation of solar- and stellar-oscillation (GONG \citep{harvey1996global} and SONG \citep{grundahl2006song}). As for future full-disk observation networks, like the anticipated Solar Physics Research Integrated Network Group (SPRING), the data homogenization between sites becomes more important, due to the improved instrumentation and automatic correction software \citep{gosain2018design}. In order to provide the highest-quality data products, an objective quality assessment is required to remove low-quality observations from the data stream as well as to compare and select between simultaneous observations from different observing sites.

The image quality provides a critical parameter for filtering observation series and frame selection \citep{popowicz2017review}. Too strong filtering may lead to to avoidable gaps in the series, while too weak filtering can reduce the quality of the data series. Automated detection methods rely on the high-quality of the input data. The quality assessment before further processing is important to guarantee the validity of the detection results \citep{potzi2018event}. Data driven methods typically improve with the size of the data set, while erroneous samples lead to a performance decrease or even failure of the method \citep{galvez2019machine}. The manual cross-checking of ten thousands of data samples is a) tedious work, b) prone to errors, and c) impossible to achieve in quasi real-time. Thus, the development of robust automated methods is essential.

Image quality metrics (IQM) have been addressed in several ways and can be categorized into three main groups according to the availability of a reference image. For full-reference IQMs a distortion free image exists, and thus the deviation from this image can be quantified. This ranges from simple pixel-based metrics, such as the mean-squared-error (MSE), to more advanced quality metrics, such as the structural similarity index (SSIM), which shows a good agreement with the human perception \citep{wang2004image}. In cases where no additional information about a reference image is available, we refer to no-reference image quality metrics (also known as blind image quality assessment). Several methods have been proposed for this problem (e.g., CORNIA \citep{ye2012cornia}, BRISQUE \citep{mittal2012brisque}). When information about a reference image is available to some extent, for instance in the form of extracted features, the quality metric is refereed to as reduced-reference IQM \citep{wang2004image}.

In case of solar observations, where it is intended to quantify the image quality for each observation, no full-reference image exists. \cite{popowicz2017review} reviewed various image quality metrics for high-resolution solar observations and provide a comparison of 36 different methods. A frequently used quality metric is the root-mean-square contrast, which has a dependence on the solar structure. More recent approaches aim to provide an objective image quality metric, such as the no-reference metric by \cite{deng2015objective}. Solar features show a strong similarity, which allows for the use of reduced-reference image quality metrics. In \cite{huang2019perception} such a metric has been proposed, based on the assumption of the multi-fractal property of solar images. 

However, for solar full-disk observations, the problem setting is different as here both global (e.g., large-scale clouds) as well as local (e.g., contrast, small clouds) effects play a role. In \cite{potzi2015real} an image quality check for full-disk H$\alpha$ filtergrams has been developed and implemented as part of the observing and data processing pipeline at Kanzelh\"ohe Observatory. The method makes use of known properties of the solar images by quantifying the deviations from a circle as fitted to the solar limb, quantifying the large-scale intensity distribution in image quadrants and estimating the image sharpness by computing the correlation with a blurred version of the original image. The weighting of these different parameters to obtain one combined image quality parameter was determined empirically, and is thus to some degree subjective. 

With the advent of deep learning methods, two important components can be taken into account, (1) the structural appearance of solar features and (2) the deviations from the true image distribution. While recent methods try to compare structural similarity over pixel-based estimations \citep{huang2019perception,deng2015objective}, to this point there exists no image quality metric for solar observations which directly estimates deviations from the true image distribution. The stability of deep learning methods relies on the variety and the quantity of training samples used. This means that high performance can be expected if new data samples are within the domain of the training set, but lacking performance for new samples that deviate from the training data. In the case of full-disk solar observations, the detection of strong deviations from regular observations is particularly important. Since strongly degraded observations are commonly removed from the data archives, a supervised approach can not fully account for the diversity of regular observation series. Therefore, we use a different approach applying unsupervised training methods.

\begin{figure*}%
\sidecaption
\includegraphics[width=\linewidth * 2/3]{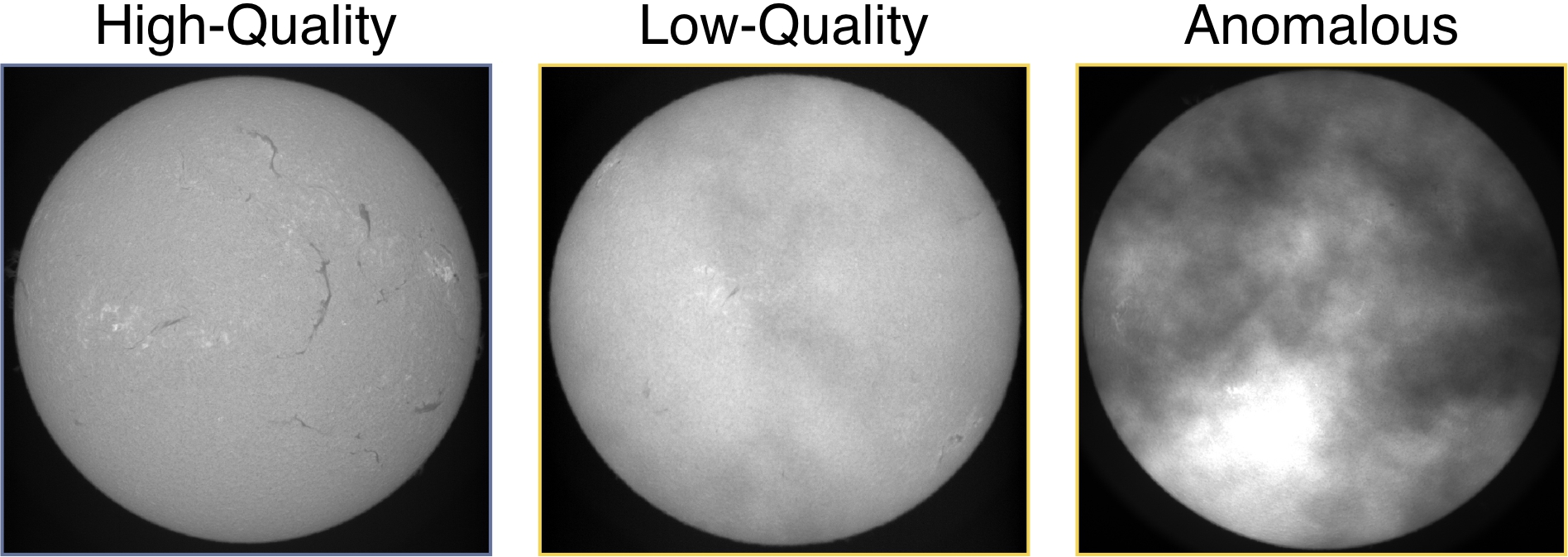}
\caption{Representative samples of the three different image quality classes. High-quality observations are characterized by sharp structures and no degrading effects (left panel). Low-quality observations suffer from degrading effects or appear blurred (middle panel). Anomalous observations show strong atmospheric influences or instrumental errors which excludes them from further scientific analysis (right panel).}
\label{figure:quality_samples}
\end{figure*}

Throughout this study, we categorize the full-disk images into three quality classes. In Fig. \ref{figure:quality_samples} a representative sample of each image quality class is shown. (1) We refer to \textbf{high-quality} observations if the image is not affected by clouds, is properly aligned to cover the full solar disk and provides the sharpness that is attainable under good observing conditions at the observing site (Fig. \ref{figure:quality_samples}, left panel). Such observations are well suited for scientific applications and for processing by automated algorithms. For the classification we do not not consider the content or scientific importance of the observation (e.g., presence of flares, filaments, active regions). We note that thin clouds often reduce atmospheric turbulence and can lead to exceptional good image quality. In this study, we only refer to atmospheric effects as clouds if they are visually recognizable in the image. (2) We refer to \textbf{low-quality} observations when a degradation in image quality can be identified. This can be induced by the turbulent atmosphere or other atmospheric effects such as thin clouds or contrails (Fig. \ref{figure:quality_samples}, middle panel). Low-quality observations can still be used for scientific analysis and visual inspection, but can lead to irregular behavior when applied to automated algorithms. (3) Observations which show strong degradation (e.g., thick clouds, partial occultation, instrumental misalignment) can typically not be used for scientific applications and are therefore removed from the data archives by existing algorithms \citep{potzi2015real}. Since these observations differ from regular observations available in the archive, we refer to them as \textbf{anomalous} (Fig. \ref{figure:quality_samples}, right panel).

In this paper, we present a novel method for no-reference image quality assessment for ground-based full-disk solar observations. We employ an unsupervised deep learning approach which uses the true image distribution of high-quality observations to detect deviations from it. Our method provides an objective image quality score and reliably detects anomalies in the data. Furthermore, our method can account for the identification of regions that are affected. The model training is performed with high-quality observations and requires no further reference image after training. In addition, we propose a classifier as an extension to our neural network architecture.  This classifier uses in addition also low-quality observations to provide an increased sensitivity for minor quality degradation. We made our codes publicly available under:
\url{https://github.com/RobertJaro/SolarImageQualityAssessment}.

\section{Data Set}
\label{section:dataset}
In this study, we demonstrate our method for solar full-disk H$\alpha$ filtergrams from Kanzelh\"ohe Observatory for Solar and Environmental Research (KSO; \url{https://kso.ac.at/}). KSO regularly takes H$\alpha$ images at a cadence of 6 seconds and provides a fully-automated data reduction and provisioning, which allows for data access in near real-time. The spatial resolution of the telescope is about 2", and the data is recorded by a 2024x2024 pixel CCD corresponding to a sampling of about 1" per pixel \citep{otruba2003halphaKSO, potzi2015real}. The quality assessment is provided by an automated algorithm which separates observations into three classes \citep{potzi2018event}. Hereby only the highest quality (class 1) is used in the pipeline of automated event detection, class 2 is still considered for scientific analysis and visual inspections, whereas the lowest quality (class 3) is completely removed from the archive. The image quality assessment criteria and division into classes were determined empirically, and are specifically adapted to the KSO H$\alpha$ filtergrams. The method has not yet been systematically evaluated. In this study, we account for this evaluation by manually classifying a test set which we compare to the existing quality assignment as well as to the newly developed deep learning approach that is presented here. 

From the KSO data archive (\url{http://cesar.kso.ac.at}), we randomly sampled solar full-disk H$\alpha$ filtergrams recorded between 2012 and 2019. Hereby we alternated between observations labeled as class 1 and class 2. We manually separated high-quality images and observations that contain clouds or blurred solar features (low-quality), until we acquired 2,000 images per quality class. Observations with strong quality degradation (anomalous) are sparse in the KSO archive and are not considered for training purposes. From this data set, we separate 1,650 observations per quality class and keep them as independent test set, which we do not use for any of our model training. The remaining 350 observations per quality class are used to automatically create the training set for the primary model (see Sect. \ref{section:dataprep}).

Observations with strong degradations (e.g., strong cloud coverage, partial eclipsed observations, overexposure) are removed from the KSO archive (class 3). In order to assert the stability for unfiltered (raw) image time series, which have a larger variety of atmospheric and instrumental effects than the training and test set, we analyze in addition 5 full observing days with varying seeing conditions (2018-09-27 till 2018-09-30 and 2019-01-26). This set has not been pre-filtered with respect to image quality. From the total 10,050 filtergrams, we manually label all images that show strong degradation. This leads to a total set of 620 images attributed to the "anomalous" class.

We note that our method is not restricted to a specific instrument, wavelength or observation target and can be applied similarly to new data sets.

\section{Method}
\label{section:method}


\begin{figure*}%
\sidecaption
\includegraphics[width=\linewidth * 2/3]{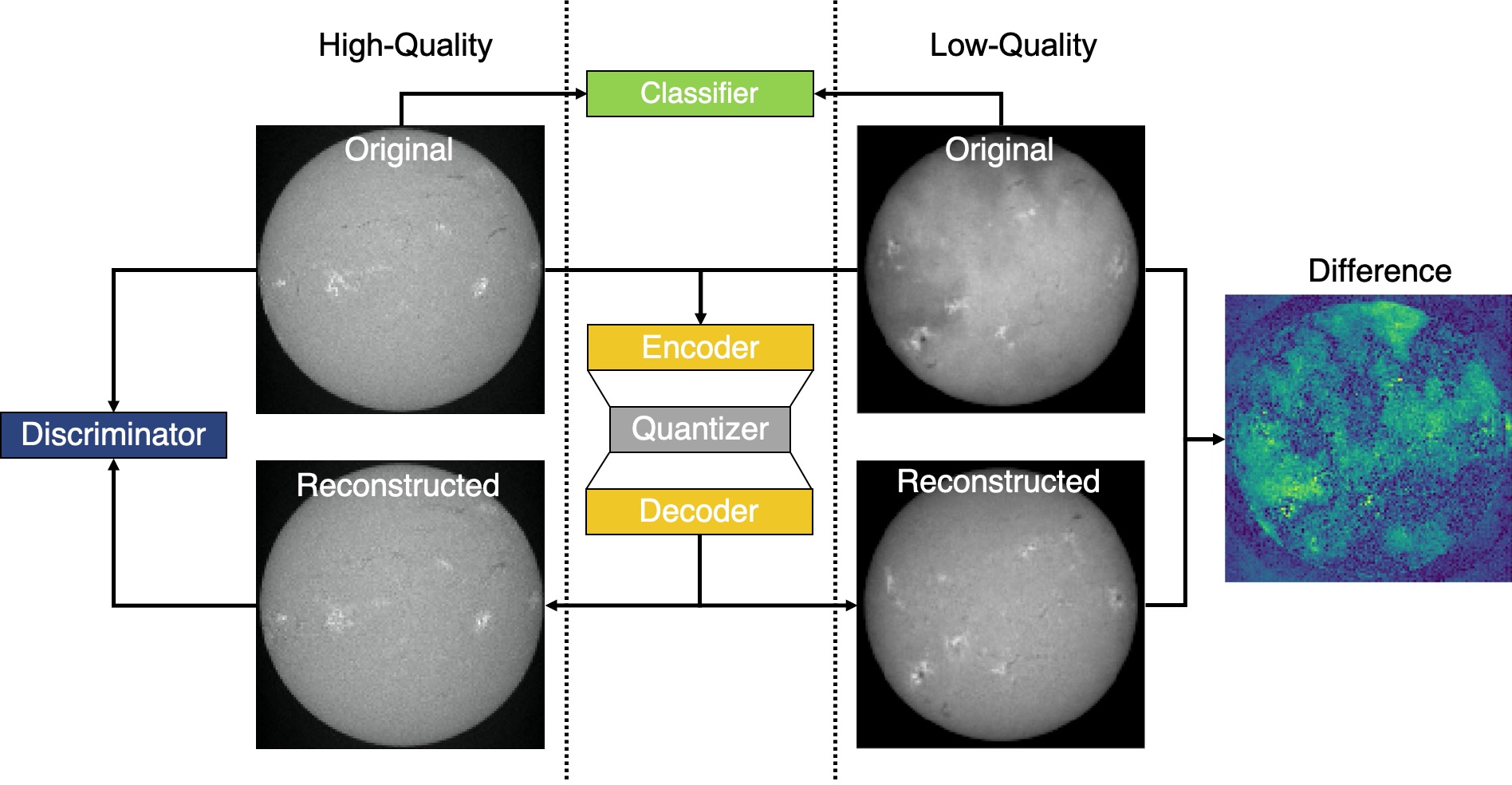}
\caption{Overview of the proposed method. The generator consists of an encoder, quantizer and decoder. The generator is trained with high-quality images (left), where the encoder transforms the original image to a compressed representation and further information is truncated by the quantizer. The decoder uses this representation to reconstruct the original image. The discriminator optimizes the perceptual quality of the reconstruction and provides the content loss for the quality metric, which encourages the generator to model the high-quality domain. In addition, an optional classifier can be used which is trained to distinguish between the two image quality classes. When low-quality images (right) are transformed by the pre-trained generator, the reconstruction shows deviations from the original, which allows to identify the affected regions and to estimate the image quality.}
\label{figure:model}
\end{figure*}

Neural Networks have shown impressive results for classification tasks \citep{he2016deep, simonyan2014very, chollet2017xception, lecun2015deep}. Applied to the case of quality classification, this could be solved as a two class problem. In this setting, a data set of high- and low-quality observations needs to be manually labeled and the network is trained to predict the correct class for a given input image (supervised training). Convolutional neural networks (CNNs) have shown the capability to directly learn from images, by automatically extracting features from edges and shapes within the image \citep{lecun2015deep}. This provides an advantage over classical machine learning approaches that provide a classification based on manually extracted features. In the present case, clouds significantly differ in shape and structure from solar features, while extracted parameters such as the global intensity distribution, primarily detect large scale deviations and are prone to falsely identify solar features. Even though a CNN can identify small changes in image quality, the classification approach has two fundamental issues in order to provide a reliable quality assessment. (1) Learning based algorithms show a high stability for data which are similar to the training set, but the coverage of all possible atmospheric and instrumental effects is not possible. Furthermore, data with strong degradations are often not stored in the archive \citep{potzi2018event}. Therefore, there is typically a large amount of high-quality observations available, while the variety of low-quality observations is sparse. For classification tasks, neural networks can produce unexpected results even for minor deviations from the training set distribution \citep{goodfellow2014explaining,papernot2017practical}. (2) The predictions of a neural network classifier are probabilistic scores and typically do not scale with the quality of the images. However, the adequate filtering of solar observations requires a proper quality metric and the identification of affected regions. The lack of information about the reasoning of the neural network is often referred to as black-box problem, which we are trying to mitigate in this study.

Instead of classification, we use an unsupervised deep learning approach to learn solely from high-quality images. Our model is composed of two main components: the encoder takes the original image as an input and compresses it into a reduced representation, while the decoder uses the encoded image to reconstruct the input image. In the encoding step, a significant amount of information is truncated. Since we are dealing with a restricted problem of a limited data set, the network can infer information about the high-quality image distribution to recover truncated information during encoding. The quality of the output is therefore determined by the amount of truncation, which is adjusted by the model architecture, and the complexity of the true image distribution. After training the network with high-quality data, it is used to identify low-quality images based on the following property of the network: When translating images of the low-quality domain, the decoder can not reconstruct the characteristics of the low-quality distribution, which leads to deviations between the original and the reconstructed image. The deviation between reconstruction and original is termed reconstruction loss.

The single optimization for a distortion metric (e.g., mean-squared-error) often shows a lack of performance and leads to blurred rather than sharp structures \citep{blau2018perceptionDistortion}.  We therefore build upon different concepts: (1) We use an adversarial loss (Sect. \ref{section:adversarial}), which allows our network to generate data, learn the characteristics of the high-quality domain and enhance the perceptual quality of the reconstruction. (2) Instead of a pixel-wise distortion metric, we optimize for feature similarity (or content loss; Sect. \ref{section:content}). This leads to a larger deviation of the reconstruction loss for low-quality images and to a better translation of solar features. As an optional component, we introduce a classifier network to our existing architecture. This network is trained in addition to high-quality images also with low-quality samples and provides a probabilistic classification into high- and low-quality data. In addition, this network is used to increase the sensitivity for low-quality observations as estimated by the content loss (Sect. \ref{section:classifier}). An overview of the combined model training is given in Fig. \ref{figure:model}.

For our primary model, we build on a multi-scale encoder-decoder architecture (similar to \cite{agustsson2019generative, johnson2016perceptual, wang2018pix2pixhd}), which has shown strong performance for image translation, style transfer and super-resolution tasks. The detailed model architecture is given in Appendix \ref{section:model_architecture}. The following sections explain the individual components of our full model setup. In Sect. \ref{section:compression}, the separation between high- and low-quality observations is discussed. Sect. \ref{section:adversarial} introduces the components to model the high-quality distribution. In Sect. \ref{section:content} the loss function for feature-based translation is introduced. An optional component to increase the model performance is the classifier network, which is described in Sect. \ref{section:classifier}. The data preparation and evaluation metrics are covered in Sect. \ref{section:dataprep} and \ref{section:metric}, respectively.

\subsection{Feature Compression}
\label{section:compression}
In order to separate between high- and low-quality observations, we aim at increasing the reconstruction loss for low-quality observations, while keeping the reconstruction of high-quality observations close to the original. We refer to the distance between the quality distributions as evaluated by the reconstruction loss as margin, which we aim to maximize. The parameter which allows for the adjustment of the reconstruction quality is the amount of truncated information by the encoder. We build upon the image compression network by \cite{agustsson2019generative} which allows for the adjustment of compression by a quantizer and provides a sufficient amount of parameters to generate images with a high perceptual quality. Hereby the separation of high- and low-quality observations is limited by an upper and lower bound of compression. If the model truncates too much information in the intermediate layers, the performance for both domains suffers and the margin becomes too small to separate the distributions. Similarly, a model which does not truncate a sufficient amount of information will reconstruct the image pixel-wise and does not learn to infer information about the true image distribution, which leads to a similar performance on both data sets. The quantizer uses the latent feature maps of the encoder and maps it to $L$ discrete levels. The information stored in the discretized representation $\hat{\omega}$ is measured by the entropy
\begin{equation}
    \label{equation:bound}
    H(\hat{\omega}) <= $ dim$(\hat{\omega})\log_2(L),
\end{equation}
which is bound by the model architecture in terms of dimensions of the feature maps $\hat{\omega}$ as provided by the encoder and number of discrete levels $L$ \citep{mentzer2018conditional, agustsson2019generative}.

\subsection{Adversarial Loss}
\label{section:adversarial}
Training neural networks with a loss based on pixel-wise differences often results in blurred images and a lack in perceptual quality \citep{isola2017image,agustsson2019generative}. This problem can be overcome with the use of generative adversarial networks (GANs), which optimize the perceptual quality based on an additional neural network. With this setup, the generation of highly realistic synthetic images is possible \citep{wang2018pix2pixhd,karras2017progressive}.

GANs as originally proposed by \cite{goodfellow2014gan} are composed of a generating network (\textit{generator}) which produces a synthetic image from a random input vector (\textit{latent space}) and a discriminating network (\textit{discriminator}) that distinguishes between generated and real images. The training is performed in a competitive setup between the generator and discriminator. In the first step, the model parameters of the generator are kept constant and the discriminator is trained to correctly classify images as either synthetic or real. In the second step, the discriminator weights are kept constant and the generator is trained to produce images which lead to a classification as real by the discriminator. In the first iterations the results are arbitrary, but from the iterative repetition of these steps both networks become experts in generating/discriminating images. In other words, the discriminator learns from the true image distribution and ensures that the generator produces images that are close to real images. By  randomly sampling inputs from the latent space, synthetic images can be produced.

We optimize the discriminator $D$ and generator $G$ for the objective proposed by \cite{mao2017least} (Least-Squares GAN):
\begin{equation}
    \label{equation:discriminator}
    \mathcal{L}_{D} = \min_{D} \mathbb{E}[\left(D(z) - 1\right)^2] + \mathbb{E}[D(G(z))^2]
\end{equation}
and
\begin{equation}
    \label{equation:generator}
    \mathcal{L}_{G} = \min_{G} \mathbb{E}[(D(G(z)) - 1)^2],
\end{equation}
where the discriminator objective $\mathcal{L}_{D}$ is given by the minimization of the expectation value of the loss as estimated by the squared difference between the discriminator prediction for the real images $D(z)$, as well as the generated images $D(G(z))$ and the assigned labels (1 for  real images and 0 for generated images). The objective of the generator $\mathcal{L}_{G}$ is obtained by minimizing the loss of the generated samples $G(z)$ for the inverted labels. In order to synthesize images, $z$ corresponds to a random input vector sampled from a prior distribution. In this way, the network learns to find a mapping between the defined prior distribution and the data distribution \citep{agustsson2019generative, goodfellow2014gan}. In the present case of image transformation, the random input vector $z$ is replaced by an image $x$ which is translated conditionally into a different domain \citep{isola2017image, wang2018pix2pixhd, karras2017progressive}. In our setup, the generating network is given by the encoder and decoder as introduced in Sect. \ref{section:method}. The encoder translates the input image into its latent space representation, while the decoder uses the encoded features to recover the original by generating the missing information from the inferred characteristics of the high-quality image distribution, as enforced by the discriminator.

The expected output of the generator $G$ is the same as the input image $x$, therefore we extend the generator loss as given in Eq. \ref{equation:generator} by a loss term that controls that the generated images $G(x)$ are close to the original $x$ (e.g., MSE). In Sect. \ref{section:content} we introduce the corresponding loss function which accounts for this term by estimating the content similarity between the original and reconstruction. We explicitly neglect an additional pixel-based loss, which would benefit the reconstruction of low-quality images. For the adversarial training objective, we follow the implementation by \cite{wang2018pix2pixhd} and use 3 discriminators which are trained in parallel and the input is rescaled by an average pooling layer with a pooling window size of 1, 2 and 4, respectively (see also \cite{agustsson2019generative}). As shown for image compression by \cite{agustsson2019generative}, the GAN approach is capable to learn common image textures and features, while methods which only use the MSE fail at modeling the true image distribution. 

For image quality assessment, we train the generator only with high-quality observations. Therefore, the network learns only to encode solar features into the latent space representation. When the trained network is used to reconstruct low-quality observations, the encoding of solar features suffers for blurred observations. Furthermore, unknown features (e.g., clouds) can not be translated into the compressed representation and will thus be misinterpreted by the encoder. Here, we make use of this property. From the deviation between the original and reconstruction, we obtain a quality metric which has a reduced sensitivity for solar features and a strong sensitivity for deviations from the high-quality distribution.

\subsection{Content Loss}
\label{section:content}
For pixel-based metrics (e.g., MSE) small shifts can cause a large increase of the reconstruction loss, which often leads to blurred results.  An alternative to the pixel-wise comparison is the evaluation of content similarity between the original and reconstructed image. This can be achieved by comparing the activation of multiple layers of a pre-trained VGG network \citep{simonyan2014very}. The network is hereby trained for a classification task, which extracts patterns at each intermediate layer. By comparing the activation of the generated and original image, a metric which is more sensitive for the content can be obtained. For our application, we define the content loss based on the discriminator, similar to \cite{wang2018pix2pixhd}:
\begin{equation}
    \label{equation:content}
    \mathcal{L}_{Content,j} = \mathbb{E} \sum_{i=0}^{4}\frac{1}{N_i}[\Vert D^{(i)}_{j}(x) - D^{(i)}_{j}(G(x))\Vert_1].
\end{equation}
$D^{(i)}_j$ refers to the layer $i$ of the discriminator $j$, $G$ to the generator and $N_j$ to the total number of features per layer. For each of our 3 discriminators we use all intermediate activation layers. Our final generator objective is given by:
\begin{equation}
    \label{equation:combined}
    \mathcal{L}_{G} = \min_{G} \sum_{j=0}^{3} \left( \mathcal{L}_{Content,j} +  \mathbb{E} [(D_j(G(x)) - 1)^2] \right).
\end{equation}
We use the content loss $\mathcal{L}_{Content,j}$ (first term) to control that the generated images are close to the original and the adversarial loss (second term) to ensure that the generated images are perceptually similar to images from the high-quality domain.

\subsection{Classifier Network}
\label{section:classifier}

In addition to the introduced architecture, we add an optional classifying network, which can be used in case that a sufficient amount of low-quality observations is available. The classifier is trained in parallel to the generator and discriminator and uses in addition also low-quality observations to provide a probabilistic prediction on the image quality. In the same way as for the discriminator, we use 3 classifiers at different resolutions. For the combined architecture, the content loss is derived from the feature activation of the classifiers instead of the discriminators. From this setup, we expect a larger margin between high- and low-quality reconstruction loss, since the classifier extracts features from both high- and low-quality images. The generator training is performed in the same way, by using only high-quality observations.

\subsection{Data Preparation}
\label{section:dataprep}

For each image we crop the frame to [-1000", 1000"], i.e. covering the full solar disk, and resize it to 128x128 pixels. This resolution is in accordance with our detection objective, where we are primarily interested in large scale degradations. We compare two different data normalizations:
\begin{enumerate}
    \item \textbf{Image Normalization:} The data is rescaled based on the minimum and maximum value to an interval of [-1, 1]. In order to reduce the impact of small-scale brightness enhancements, we crop values outside 3$\sigma$ from the mean of the image prior to normalizing.
    \item \textbf{Contrast Normalization:} For each image we subtract the median and divide the result by the standard deviation of the image. Values outside [-2.5, 2.5] are cropped and afterwards rescaled linearly to [-1, 1] \citep{goodfellow2016deep}.
\end{enumerate}
The contrast normalization centers the data to the mean, which makes the trained network more sensitive to shifts in the image intensity distribution (e.g., induced by partially occulting clouds). For the image normalization, we found a better accordance with the human perception when identifying clouds. A normalization based on the adjustment for exposure time and normalizing to a fixed value range provides less correlation with the apparent image quality. This is primarily due to the dynamic exposure time which compensates low intensity observations (e.g., caused by clouds) and high intensities (e.g., ongoing flares).

Deep Learning methods benefit from a larger amount of training samples, therefore we automatically extend our manually labeled data set. We assume that image classification can be used to detect most quality degrading effects and train the classifier network introduced in Sect. \ref{section:classifier} for a basic classification task. We use the 700 annotated observations which were not considered for the test set (Sect. \ref{section:dataset}) and apply the same parameter configuration as for the classifier training as given in Sect. \ref{section:result}. The trained model is used to automatically annotate a new data set of 20,184 high-quality and 2,198 low-quality observations. This potentially introduces more misclassified samples in the training set of the primary model, but we found that a larger data set improves the performance and stability of the more challenging translation task.

During model training, we separate 10\% of the training set for validation purposes. We note that we do not apply a strict temporal separation of the training and test set, since we are mostly interested in short term variations. We found that the use of a large random data set is sufficient.  Table \ref{table:data_set} provides a summary of the considered data sets.

\begin{table}
\caption{Data sets for training and evaluation of the proposed models. "Manual" indicates that a data set was labeled by a human. The number of samples include both classes (high- and low-quality). The number in brackets refers to the fraction of manually labeled samples. Validation sets are split from the training sets on demand.}             
\label{table:data_set}      
\centering                          
\begin{tabular}{c || c c c}        
\hline\hline                 
Data Set & Manual & Samples \\    
\hline                        
   General Test & \cmark & 3,300 \\ 
   Classification Training & \cmark & 700 \\ 
   General Training & \xmark & 22,382 \\
   KSO Unfiltered Test & partially & 10,050 (620) \\
\hline                                   
\end{tabular}
\end{table}

\subsection{Quality and Evaluation Metric}
\label{section:metric}

For the evaluation of the reconstruction quality we use four different metrics:
\begin{enumerate}
    \item \textbf{Mean-Squared-Error (MSE):} provides a pixel-wise loss which gives a good estimate for larger regions, but suffers from small-scale differences.
    \item \textbf{Content-Loss (Sect. \ref{section:content}):} provides a metric optimized for the considered data set. This metric compares image features over pixel-wise differences.
    \item \textbf{Structural-Similarity-Index (SSIM):} provides a good correspondence with the human vision and is based on image similarity \citep{wang2004image}.
    \item \textbf{Classification (Sect. \ref{section:classifier}):} gives a probabilistic prediction on the image quality, but does not provide a continuous metric and requires a manually annotated data set with low-quality images. The classification can better account for minor atmospheric effects than the continuous metrics, but might lead to wrong predictions for strong deviations from the training set.
\end{enumerate}

The metrics resemble the image quality, where larger losses indicate a stronger deviation from the original and therefore a lower quality. In order to separate the image into high- and low-quality, we apply thresholds according to the evaluation of the validation set. The content-loss shows the largest margin between the individual classes and is therefore considered as our primary quality measure. We use the MSE and SSIM for additional verification. The result of the classifier can not account for a quality measure and can lead to unexpected behavior for anomalous data. Therefore, for our classification scheme we combine the classifier predictions with the content loss.

The combined classifier is composed of three networks each predicting patch-wise at a different resolution, with an output of an 8x8 grid. We take the mean result per patch and sum over the classifiers. We classify images as low-quality above a threshold of 1, which corresponds to the classification as low-quality image by at least one classifier.

In addition, we identify anomalies by the continuous quality metric. To this aim, we use the content loss and scale it according to the results on the validation set. We define the low-quality threshold at 3$\sigma$ above the mean and scale the data between 0 and 4 times the low-quality threshold. From this scaling, observations with quality 0 refer to a perfect reconstruction, 0.25 defines the low-quality threshold and values above 1 correspond to observations with strong degrading effects (anomalous observations). For the base architecture without a classifier, we identify anomalies solely on the content loss. Hereby, we lower the threshold to 2$\sigma$ above the mean of the high-quality distribution as evaluated on the validation set and leave the threshold for anomalous observations unmodified.

We evaluate the correct predictions of low- and high-quality images in terms of accuracy and the True-Skill-Statistic (TSS; also known as Hanssen \& Kuipers Discriminant) \citep{ barnes2016comparison, potzi2018event}
\begin{equation}
    \label{equation:kss}
    TSS = \frac{TP}{TP + FN} - \frac{FP}{FP + TN}.
\end{equation}
The variables correspond to the entries of the confusion matrix as number of true positives (TP), true negatives (TN), false positives (FP) and false negatives (FN).

\section{Results}
\label{section:result}
For the model training we apply different parameter settings and evaluate them according to the metrics introduced in Sect. \ref{section:metric}. We use the short hand notation of (CLASS/DISC)-qX-(IMG/CONTR) where DISC refers to the base architecture and CLASS to the classifier extension. X denotes the number of channels in the compressed representation. All our models use 5 discrete levels for quantization. We compare both data normalizations, where IMG refers to the image normalization and CONTR to the contrast normalization. For example, CLASS-q8-IMG refers to the network extended by a classifier, with 8 filters in the quantizer and normalizes each image by the range between its minimum and maximum value.

We train each of our models for 300.000 iterations until the MSE and content loss of the low-quality samples in our validation set start to converge towards an upper bound. We use the Adam optimizer with a learning rate of 0.0002 and set $\beta_1 = 0.5$ and $\beta_2 = 0.9$ \citep{kingma2014adam}.

\begin{table*}
\caption{Performance of the different model settings, as evaluated on the manually classified test set. Content, MSE and SSIM refers to the margin between the median losses of the two quality distributions for the according metric. Compression refers to the number of channels used for the quantized representation. Correct classifications are determined by the criteria introduced in Sect. \ref{section:metric} and are quantified in terms of accuracy and TSS. The evaluation of the original KSO labels is given in the last row. The results of the model yielding the best performance (CLASS-q8-CONTR) are marked in bold face.}             
\label{table:evaluation}      
\centering                          
\begin{tabular}{c c l || c c c c c c}        
\hline\hline                 
Classifier & Compression & Data Normalization & High-Quality MSE & Content & MSE & SSIM & TSS & Accuracy \\    
\hline                        
   \cmark & 8 & ImageNorm & 0.0044 & 0.37 & 0.013 & 0.03 & 0.95 & 97.7\% \\
   \cmark & 8 & ContrastNorm& \textbf{0.0017} & \textbf{0.72} & \textbf{0.042} & \textbf{0.08} & \textbf{0.97} & \textbf{98.5\%} \\
   \cmark & 1 & ImageNorm& 0.0044 & 0.37 & 0.006 & 0.02 & 0.95 & 97.7\% \\
   \cmark & 1 & ContrastNorm& 0.0019 & 0.45 & 0.003 & 0.01 & 0.95 & 97.7\% \\
   \xmark & 8 & ImageNorm& 0.0038 & 0.12 & 0.005 & 0.02 & 0.82 & 89.3\% \\
   \xmark & 8 & ContrastNorm& 0.0016 & 0.18 & 0.004 & 0.04 & 0.92 & 96.1\% \\
   \xmark & 1 & ImageNorm& 0.0042 & 0.18 & 0.006 & 0.03 & 0.93 & 96.5\% \\
   \xmark & 1 & ContrastNorm& 0.0018 & 0.47 & 0.028 & 0.09 & 0.93 & 96.5\% \\
\hline                                   
 	\multicolumn{3}{c||}{Baseline KSO Quality Classification} & - & - & - & - & 0.30 & 64.2\%   \\
 \hline
\end{tabular}
\end{table*}

\subsection{Observation Quality Metric}

For each of our models, we evaluate the performance metrics on the manually labeled test set (3,300 samples). The model parameters and results are summarized in Table \ref{table:evaluation}. We vary the architecture, the number of compression channels and the data normalization. The average error in the reconstruction, as measured by the MSE of the high-quality images, serves as a performance indicator for configurations with the same normalization. The ability to separate between high- and low-quality observations is estimated by the margin between the mean of the high- and low-quality distribution for each of our metrics. We evaluate the margin in terms of content loss, MSE and SSIM. From the thresholds defined in Sect. \ref{section:metric} we estimate the accuracy and TSS.

The CLASS-q8-CONTR model shows the best performance for each metric despite the SSIM. With an average content loss margin of 0.72, an accuracy of 98.5\% and a TSS of 0.97 it clearly provides the best result, while the other classifier configurations are similar in performance in terms of accuracy (cf. Table \ref{table:evaluation}). The DISC-q8-IMG model shows a significantly lower performance, as compared to the other configurations, which all achieve accuracy scores above 96\%. As can be seen from the accuracy and TSS, the number of compression channels and normalization has little impact on the ability to separate between high- and low-quality images. In contrast, the classifier architecture results in a performance increase by at least 1.2\%. The accuracy of the classifier without a low-quality threshold is 96.9\% and 98.1\% for the ImageNorm and ContrastNorm, respectively. This is an improvement by about 1\% for the models with image normalization, while for the top performing network the additional quality threshold has only a minor effect (+ 0.4\%) . The CLASS-q1-CONTR model has a lower accuracy as compared to the classifier prediction (-0.4\%).

Using the same data set, we compare the results of our deep learning algorithm also with the empirical KSO image quality assessment method in \cite{potzi2015real}. The accuracy and TSS of the empirical method, listed in the bottom line of Table \ref{table:evaluation}, have values of 0.3 and 64.2\%, respectively. This is significantly lower then the new algorithm presented here. We further randomly sampled observations from the full KSO archive and compared the original KSO labels with the deep learning model predictions, showing an agreement of 72.5\% between the two methods.

\begin{figure*}
\includegraphics[width=\textwidth]{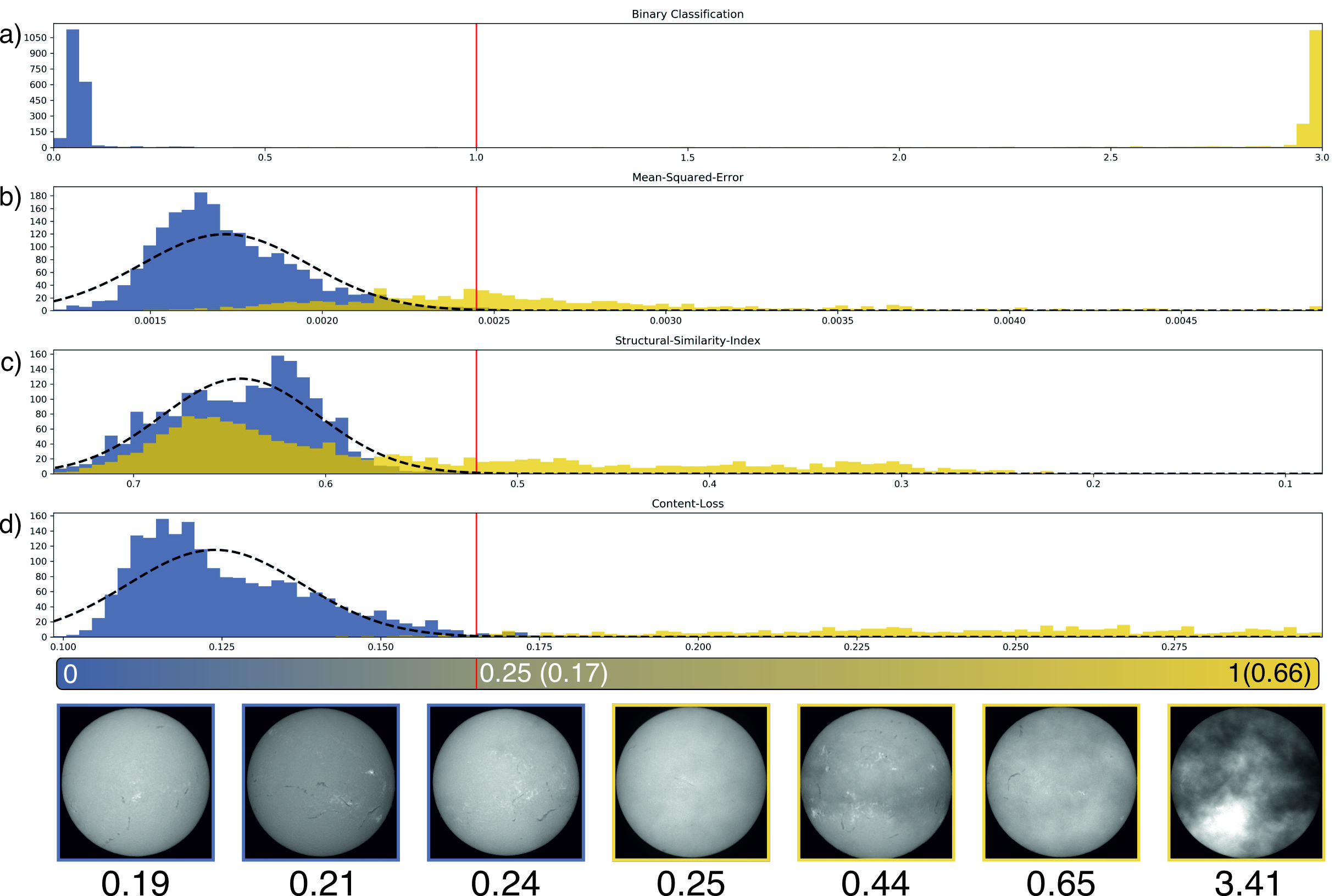}
\caption{Evaluation of the test set for our CLASS-q8-CONTR model. High-quality and low-quality samples are shown in blue and yellow, respectively. a) Distribution of the classifier predictions, b) - d) image quality metrics between the original and reconstructed image in terms of b) Mean-Squared-Error, c) Structural Similarity and d) Content Loss. The normal distribution (dashed black lines) of the high-quality images as well as the low-quality threshold (red line) is indicated for each metric. Samples of decreasing quality as evaluated by our metric and their corresponding content loss are shown in the bottom panels. The image outline is set according to the classifier prediction. An animation of the full test set with increasing image quality can be found in the supplementary material (\url{https://youtu.be/9MvdLDtxKBo}).}
\label{figure:evaluation}
\end{figure*}

Fig. \ref{figure:evaluation} visualizes the test set evaluation of our top performing network (CLASS-q8-CONTR). Panels a)-d) show the evaluation of the manually labeled test set on the defined quality metrics. High-quality samples are marked in blue and low-quality samples in yellow. The red line indicates the low-quality threshold as obtained from the evaluation of the validation set. The plots are centered to the high-quality distribution; quality estimates outside the given range are not included. Panel a) shows the evaluation of the classifier, where a large separation between the high- and low-quality class distributions can be seen, with almost no overlap at the low-quality threshold. This is in agreement with our assumption that a simple classification approach can detect most quality degrading effects. The quality metrics in b)-d) are based on the difference between the original and reconstructed image and  show a continuous transition between the two classes. The MSE in panel a) provides a distinct separation of the two distributions, while an even larger margin is achieved for the content loss (panel d). The SSIM in panel c) shows the weakest performance in separating the two classes, where only a fraction of the low-quality observations show a distinct deviation from the high-quality distribution.  The quality value is scaled as discussed in Sect. \ref{section:metric} and is indicated by the horizontal bar at the bottom of Fig. \ref{figure:evaluation}. Hereby, the values in brackets indicate the content loss at the given thresholds. From the test set we randomly selected quality estimates across the scale. The samples are shown at the bottom of Fig. \ref{figure:evaluation}.  The low-quality threshold is given at 0.25 and images with a quality estimate above 1 are considered to suffer from strong atmospheric degradation or instrumental errors. 

We found that the off-limb region can cause high quality scores, therefore we apply an on-disk correction. Especially faint clouds across the full disk can severely impact the reconstruction at the solar limb, which is not in correspondence with the image quality of the solar disk. For that reason, we remove the off-limb region before evaluating the content loss and apply an offset to align the adjusted scale with the original quality measure of the high-quality samples.

We note that the definition of the threshold was selected to suite most applications, but can be adapted for specific demands (e.g., selection of very high quality observations). A video that visualizes quality samples over the full scale can be found in the supplementary material (\url{https://youtu.be/9MvdLDtxKBo}).

\subsection{Region Identification}
\label{section:region}

\begin{figure}%
\includegraphics[width=\linewidth]{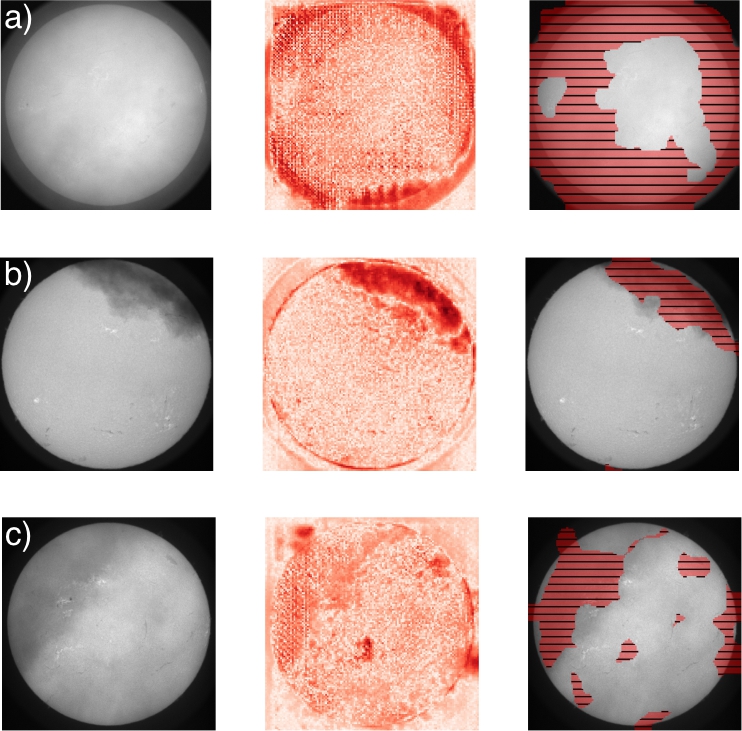}
\caption{Identification of affected regions based on the difference between the original and reconstructed image. The first column shows the input image. In the second column the difference between the original and reconstructed image is plotted on a square-root scale. Column 3 shows the original image with an overlay of the identified regions.}
\label{figure:masks}
\end{figure}

In addition to the quality metric, the reconstructed image is used to identify the affected regions within the image. CNNs show a relation between the spatial position of features in the image and the activation within the network \citep{zhou2016cnnfeatures}. Based on this property of the network, we may assume that local atmospheric effects in the original image can only cause deviations in a certain region of the reconstruction. From the difference between the original and reconstruction, regions with degrading effects can be detected. Fig. \ref{figure:masks} shows three examples of low-quality observations and the regions identified by the the CLASS-q8-IMG model. For a first representation, we use the absolute difference map between the original and reconstructed image and visualize it on a square-root intensity scale (column 2 in Fig. \ref{figure:masks}). In order to obtain the regions affected by strong degradations, we smooth the difference map with a total variation filter with a weight of 0.2 \citep{chambolle2004algorithm} and apply a threshold of 0.1, corresponding to the upper limit of the low-quality classification (column 3 in Fig. \ref{figure:masks}).

\begin{figure*}%
\sidecaption
\includegraphics[width=\linewidth * 2/3]{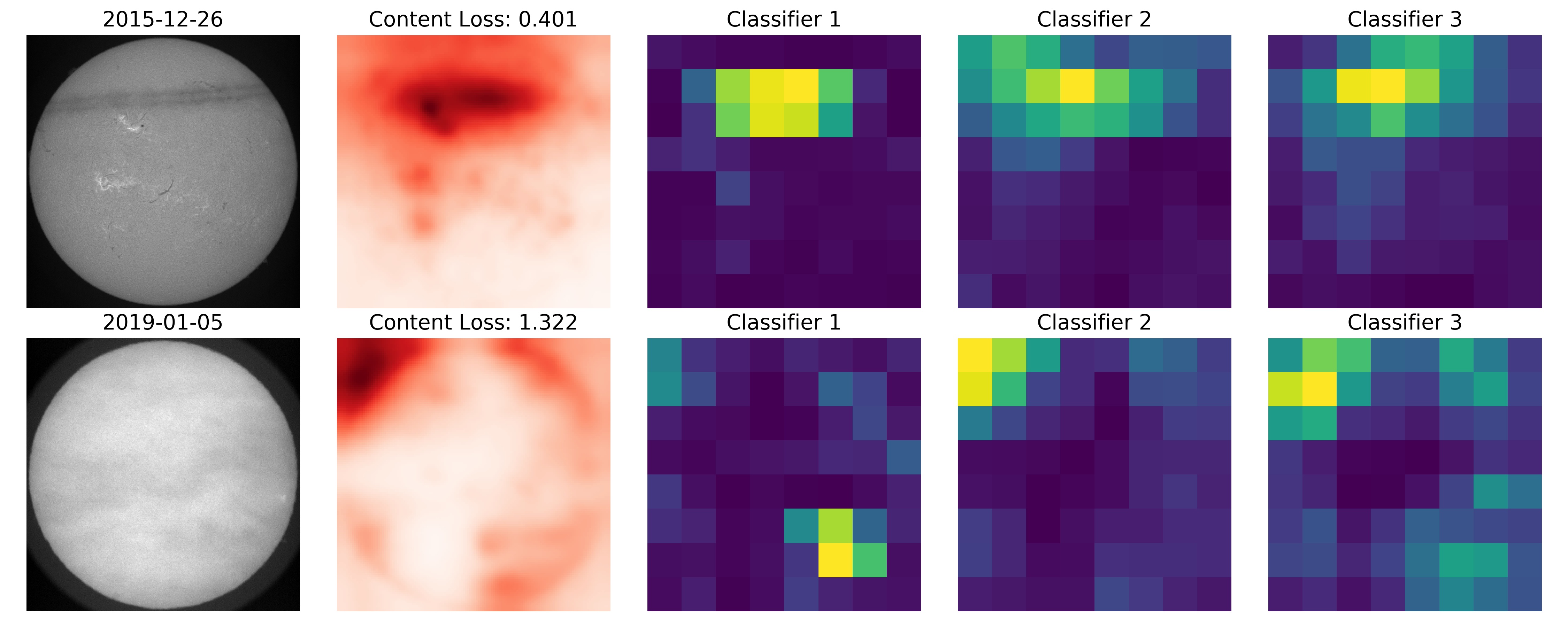}
\caption{Two examples of the region identification based on the content loss. The first column shows the input image. Column 3-5 represent the absolute differences between the feature activation of the original and reconstructed image for the three classifiers. The feature maps shown were taken at a resolution of 8x8 pixels. Column 2 shows the averaged difference over all feature maps, and lists the corresponding value of the content loss.}
\label{figure:content_masks}
\end{figure*}

Our quality metric is optimized for feature similarity and since features do not necessarily align pixel-wise with the reconstruction, we define a region identification based on the content similarity. To this aim, we utilize the same networks as for the content loss to obtain the feature activation at each layer, depending on the architecture either the classifier or discriminator network. From the fully convolutional architecture of the networks, a regional correlation between the feature activation and the position within the image can be drawn \citep{zhou2016cnnfeatures}. From each discriminator/classifier we extract the feature activation of the original and reconstruction, compute the absolute difference at each resolution level and compute the mean over the channels. We compute the region map by upsampling the individual difference maps to the original image size and summing pixel-wise over all maps. Fig. \ref{figure:content_masks} shows the resulting difference maps for two examples and the mean absolute feature differences at a resolution of 8x8 pixel for each of the classifiers.

\subsection{Application to Unfiltered Time Series}
\label{section:series}
\begin{figure*}%
\includegraphics[width=\linewidth]{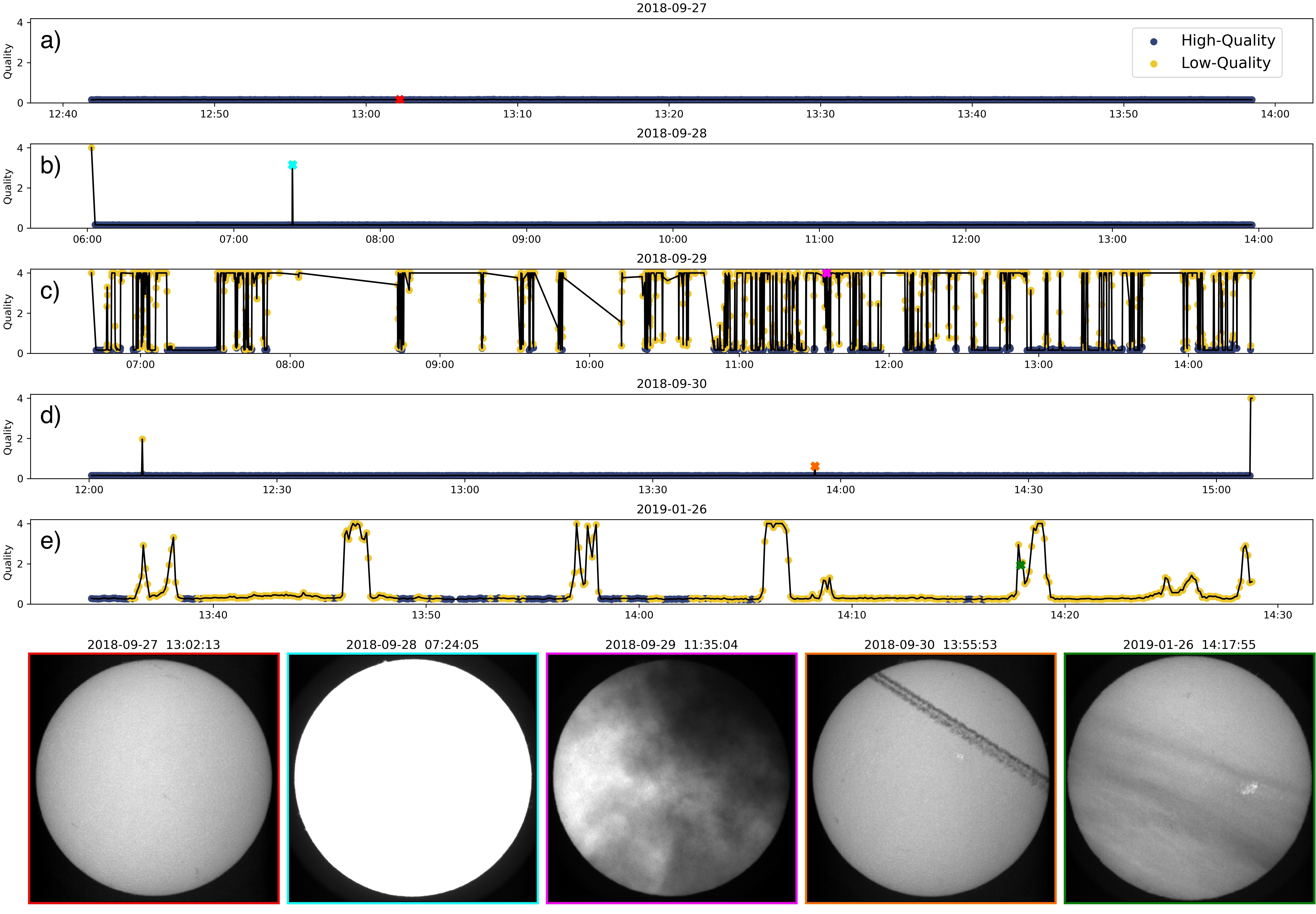}
\caption{Overview of five unfiltered observing days of KSO H$\alpha$ full-disk imaging as evaluated with the CLASS-q8-CONTR model. Panels a) to e) show the derived image quality as function of time. Examples across the series are given at the bottom. The first example shows a clear observation, while the subsequent examples show overexposure, strong cloud coverage, a contrail and partial cloud coverage. Panel a) shows a day of clear observing conditions with no quality degradations. b) has in general high quality conditions with a few overexposed images. In c) a varying cloud coverage leads to frequently changing quality scores throughout the day. d) has in general high quality conditions but a few images reveal degradation due to contrails.  e) shows a general quality decrease by clouds, while the quality further gradually decreases when denser clouds transition the disk. Videos for each observing day are available in the supplementary material.}
\label{figure:series}
\end{figure*}

In the regular observation mode at KSO, images of very low quality are rejected from further scientific use and are in general also automatically removed from the archive. In order to estimate the stability of our method to identify even strong deviations from regular observations, we use unfiltered data series of five observing days with varying observing and seeing conditions. From the full series, 620 samples with strong quality degradation are manually labeled as described in Sect. \ref{section:dataset}. From the predictions of the CLASS-q8-CONTR model we found that all samples exceed the low-quality threshold, both in terms of the binary classification and the quality threshold. From our quality scaling, we expect anomalous observations to have a quality value greater than 1. Out of the 620 samples, 617 exceed this threshold, which corresponds to a 99.5\% accuracy in terms of identification of anomalous observations for the observing days studied. Fig. \ref{figure:series} gives an overview of the individual days as evaluated with the CLASS-q8-CONTR model. The panels show the individual observing days. The series is characterized by smooth transitions for gradual variations in image quality as well as sharp jumps for sudden anomalies (e.g., appearance of clouds or contrails). At the bottom of Fig. \ref{figure:series}, examples for various effects are shown (i.e., varying cloud coverage, overexposure, contrails). We note that the overexposed image at the bottom of Fig. \ref{figure:series} is a special observing mode at KSO, used to enhance faint off-limb structures like prominences above the limb. The full per-frame overview as a video is provided in the supplements (\url{https://www.youtube.com/watch?v=sCKDFREpJEw}).

\section{Discussion}
\label{section:discussion}
The primary aim of our method is to derive a continuous quality estimator which correlates with the human perception and allows for the reliable identification of distorted solar full-disk images. This is accomplished by quantifying the deviation from the high-quality image distribution and by the use of a measure based on feature similarity. The performance of our model configurations is estimated by the ability to separate observations into high- and low-quality classes. Hereby, the manually labeled test set gives an independent measure for the agreement between the human estimate and the model classification.

As can be seen in Table \ref{table:evaluation}, the margin between the high- and low-quality distribution directly correlates with the model performance in terms of accuracy and TSS. The content loss, MSE and SSIM margin are in basic agreement with each other, while the content-loss shows the best separation between the two distributions, as can be seen from the evaluation of the test set in Fig. \ref{figure:evaluation} b) - d). All metrics are characterized by a smooth transition between quality classes, with a larger spread of the low-quality distribution. This is in agreement with the samples of the test set, which include a broad range of different types of atmospheric effects. The samples in Fig. \ref{figure:evaluation}, as well as the video containing the evaluation of the full test set (included in the supplementary material), show a good agreement with the assigned quality scores.

\subsection{Model Performance}

For all 8 model configurations, we found an overall high performance. Our best performing network (CLASS-q8-CONTR) achieves an accuracy of 98.5\% and a TSS of 0.97 in separating high- and low-quality images. The other configurations provide similar performance  with accuracies smaller by about 1-2\%. Only the DISC-q8-IMG model shows much lower performance with an accuracy of 89.3\%. We note that this is likely due to differences in the training process. With only one significant deviation out of 8 trained models, we conclude that our approach achieves stable high performance for various configurations. The performance obtained by these models is significantly higher than the present empirical algorithm that is used in the observing pipeline at Kanzelh\"ohe Observatory (described in \cite{potzi2015real}), for which we obtained an accuracy of 64.2\% and a TSS of 0.30 for the same test set.

In the case of available low-quality observations, the classifier architecture can boost the model performance by about 1-2\% accuracy and adds additional robustness to the model predictions. This can be seen from the classifier configurations, which provide an accuracy of at least 97.7\%, even though the performance varies in terms of content loss margin (Table \ref{table:evaluation}). The classifier provides a probabilistic quality-class assignment, which requires the combination with the content loss to account for a continuous quality measure. The low-quality threshold applied for classification has only a minor effect on the performance in terms of accuracy, but shows an increase for most models by 0.4-0.8\%.

The different normalization and compression channels show similar results and mostly affect the quality threshold, which has a stronger impact on the architectures without a classifier. Especially at the high- to low-quality threshold, the class assignment becomes more subjective, which leads to expected deviations in model performance.

While the contrast normalization provides a better performance, the specific choice of normalization reveals a low impact on the overall result. The choice of normalization becomes more important for the identification of clouds in the image. From a visual inspection, the image normalization provides a better correspondence to regions covered by dense clouds, while the contrast normalization is centered to the mean and normalized by the standard deviation, which leads to stronger shifts in the intensity distribution by clouds and tends to identify overexposed regions. 

\subsection{Region Identification}

The identification of clouds and other quality decreasing effects provides additional information to the quality metric. As can be seen from Figs. \ref{figure:masks} and \ref{figure:content_masks}, the reconstruction loss aligns with the affected regions in the original image. The network correctly reflects the impact of localized clouds (Fig. \ref{figure:masks}b) and global quality degrading effects (Fig. \ref{figure:masks}a).  Due to the dynamic exposure time, observations can show faint regions covered by clouds and overexposed regions simultaneously (Fig. \ref{figure:masks}c), which is detected by the neural network as deviation from an expected mean intensity value. The model is trained for content similarity, therefore the reconstruction does not align pixel-wise with the original. This induces a sensitivity for solar features in the difference masks, as can be seen from Fig. \ref{figure:masks}. While this can be mitigated with the use of extracted features from the discriminator or classifier (Fig. \ref{figure:content_masks}), the detection can only provide a coarse localization.  The generated masks based on the content loss are produced by averaging all feature activation differences,  which causes a suppression of small deviations, as can be seen in the second row of Fig. \ref{figure:content_masks}.

\subsection{Stability and Training}

Neural networks are often considered as black box, owing to the fact that it is difficult to extract information on the reasoning within the network. Here we are examining the model outputs and training progress, in order to obtain information on the functionality and stability of our approach.

From the regions obtained by the difference between the reconstruction and original image in Sect. \ref{section:region} it can be seen that deviations in reconstruction are spatially aligned with the regions of reduced quality in the original image. The extractions at different layers within the network (Fig. \ref{figure:content_masks}) reveal that the main contribution to the content loss is due to quality degradation, rather than solar features. From columns 3-5 in Fig. \ref{figure:content_masks} it can be seen that each classifier is capable of extracting different features. This finding suggests that our metric provides an objective quality assessment, which covers multiple scales, includes multiple image features and provides an enhanced sensitivity for atmospheric effects.

Neural networks can reveal large changes in the prediction, even for minor changes to the input \citep{goodfellow2014explaining, papernot2017practical}. Our model successfully detected 99.5\% of the anomalous observations in the unfiltered time series (Sect. \ref{section:series}), which proves the robustness of the chosen approach. The image quality of the unfiltered series correctly reflects the smooth transition on decreasing image quality and also captures the sudden appearance of clouds. As can be seen from days with generally good observing conditions (Fig. \ref{figure:series}a,b,d) the model shows a high stability over the time series. The increased image quality value in Fig. \ref{figure:series}e is in correspondence with the poor observing conditions at this day. We conclude that the model is not prone to small deviations, as can be commonly observed for neural networks that are applied to classification tasks \citep{goodfellow2014explaining,papernot2017practical}.

An important component for the success of our method is the truncation of information during encoding, which increases the margin between the distribution of high- and low-quality observations as evaluated by the proposed distortion metrics (Sect. \ref{section:metric}). This is controlled by the amount of channels in the quantizer. The architecture with 8 compression channels reduces the information to approximately 1\% of the original input. We found that for lower compression rates the model falls back to a pixel-wise reconstruction, which decreases the sensitivity for faint clouds. Contrarily a higher compression can reduce the reconstruction capability, which results in an increased sensitivity to the intrinsic solar features. This can also be seen from Table \ref{table:evaluation}, where the models with 8 compression channels show a lower average high-quality MSE than the models with the same normalization and 1 compression channel. Table \ref{table:evaluation} also shows that a better reconstruction of the image is not necessarily beneficial for the identification of low-quality images. This can be seen from the DISC-q8-IMG model, which achieved the lowest accuracy (89.3\%), while producing the best reconstructions (0.0038 high-quality MSE) among the models that use image normalization. 

\begin{figure}%
\includegraphics[width=\linewidth]{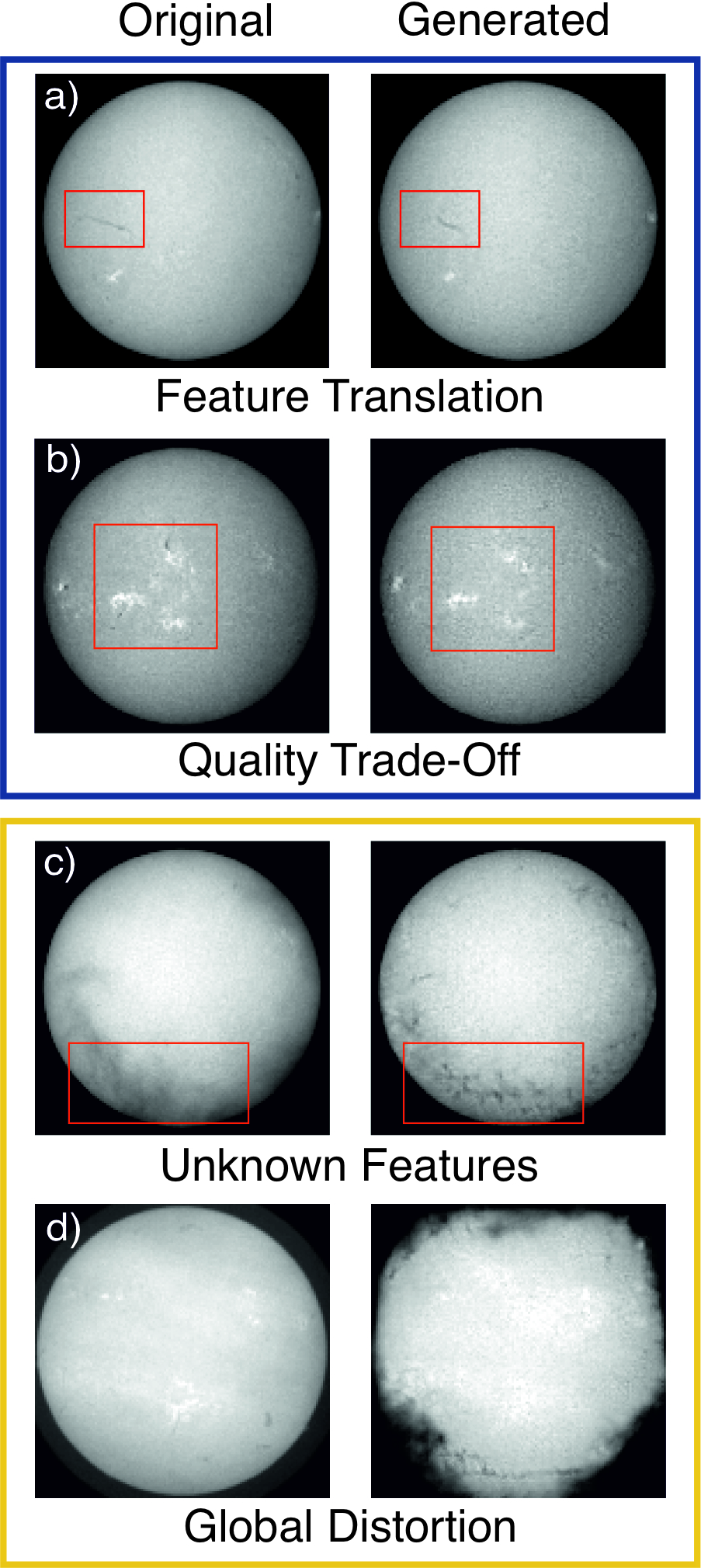}
\caption{Pairs of original H$\alpha$ filtergrams (left) as provided to the model and the resulting reconstruction (right). The top panel shows samples of the high-quality distribution, while the bottom panel contains samples from the low-quality domain. a) Illustration of the feature based image translation. The reconstructed filament can be clearly identified, but shows differences in appearance from the original. b) Example of the  image quality decrease of an high-quality image due to the strong compression. c) For low-quality images, the unknown features result in artifacts in the reconstruction. d) Global atmospheric effects show strong differences in the reconstructed image.}
\label{figure:comparison}
\end{figure}

As illustrated in the upper row of Fig. \ref{figure:comparison} the model is capable to reconstruct high-quality observations to be close to the original image. As a result of the training with the content loss, the reconstructions show a feature-based translation rather than a pixel-based. An example of this behavior is shown in  Fig. \ref{figure:comparison}a, where the reconstructed filament can be clearly identified, but appears differently in shape. The feature-based reconstruction becomes more evident for low-quality images, where the network fails to translate unknown features. In Fig. \ref{figure:comparison}c a low-quality image with clouds partially occulting the solar disk  is shown. As a result, the network reconstruction shows a strong deviation of the original dark structure and produces a structure with filament like appearance. For global atmospheric effects (i.e., coverage by faint clouds) the reconstruction yields even stronger deviations, as can be seen from Fig. \ref{figure:comparison}d. While the strong compression increases the sensitivity for unknown features, it leads to a trade-off in reconstruction quality for high-quality images (see Fig. \ref{figure:comparison}b).

\begin{figure}%
\includegraphics[width=\linewidth]{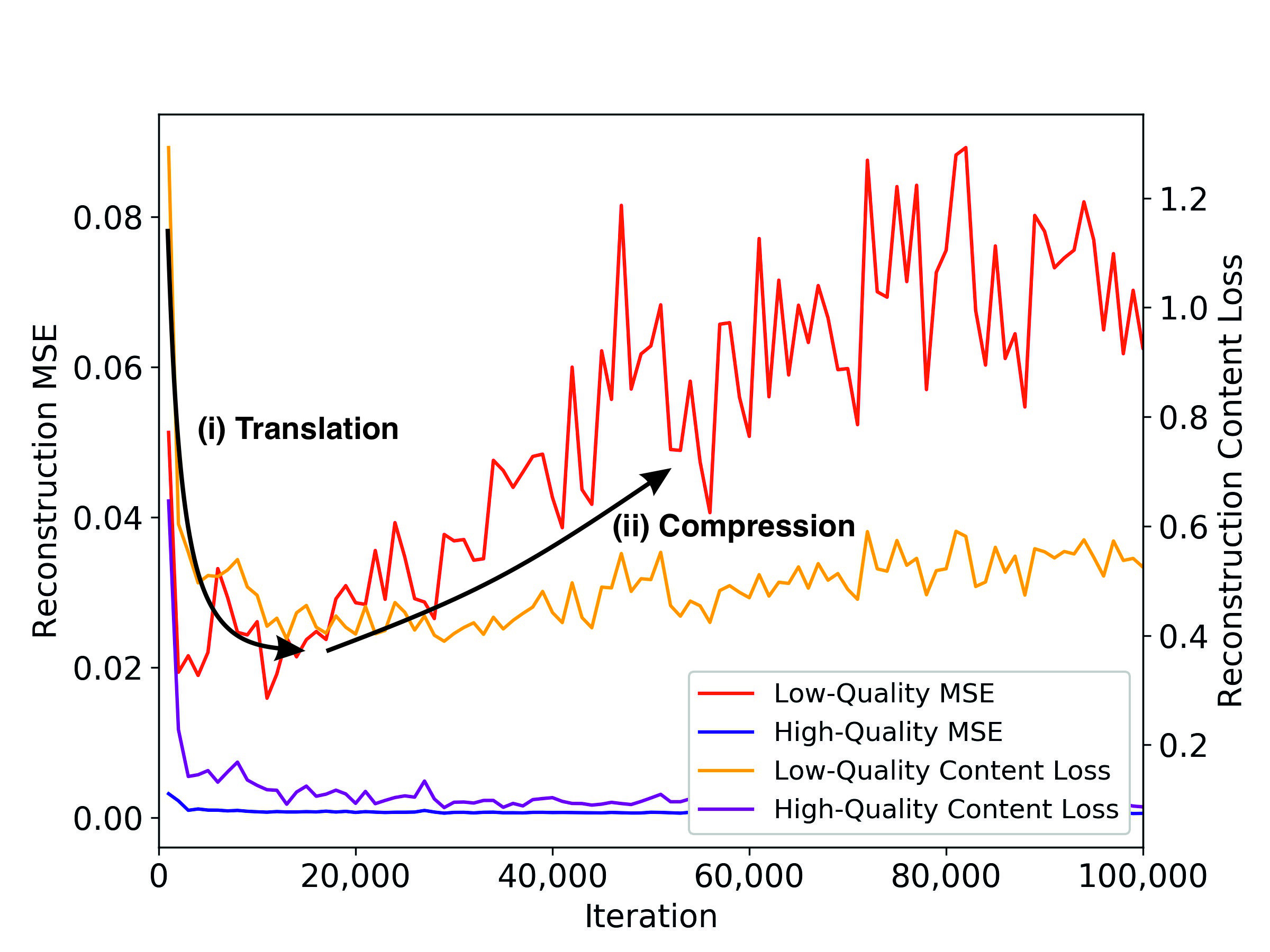}
\caption{Evaluation of the validation set during the training progress. From both quality metrics (MSE and content loss), two distinct phases can be identified. The orange and red line denote the reconstruction performance of the low-quality distribution, while the blue and purple lines are the reconstruction losses of the high-quality distribution.}
\label{figure:learning}
\end{figure}

From the evaluation of the validation set we identify two characteristic phases during training, which support our assumption of a feature based translation (Fig. \ref{figure:learning}). (i) The translation phase, where the network learns to reconstruct the original image by preserving a maximum amount of information in the quantizing layer, which increases the similarity between the original and reconstructed image for both high- and low-quality images. (ii) The compression phase, where the network starts to learn from the image distribution and truncates information during encoding, which improves the reconstruction of high-quality images, while low-quality images suffer from the learned feature compression.

\subsection{Applicability}

Our model requires approximately 17 ms for a single image quality estimation using a Nvidia Tesla M40 GPU. For a CPU based prediction, the model takes approximately 93 ms per observation when using 15 cores. In both cases, we assumed that the observations are already loaded and prepared as described in Sect. \ref{section:dataset}. Compared to the observing cadence of 6 s of the KSO H$\alpha$ filtergrams, our image quality assessment thus requires just a minor fraction of the total acquisition time. This allows for the real time application of our method, even for limited computational resources. For higher cadence modes, the performance can be linearly increased by parallelizing the computations.

 This study presents a first approach to image quality assessment for solar full-disk observations. The application to different telescopes, different filters or even multiple filters can be easily accomplished. This involves the labeling of a new data set and training the neural network as proposed by our method. A sufficiently large data set only requires a few hundred coarsely labeled images (binary classification). In case of sparse low-quality observations available, the basic architecture can be considered, which only requires high-quality observations. The presented quality assessment for KSO observations is only valid for full-disk images, this is especially important for instrumental misalignment, where a partial disk would result in a large reconstruction loss. In order to assess the quality of smaller regions, the network can be trained with image patches, which omits the encoding of the full-disk. Future developments of this method offer great potential for related applications. (1) \textbf{High-resolution solar observations} rely on identifying the best observations for post-facto algorithms (e.g., speckle-interferometry \citep{popowicz2017review}). The capability of our method to learn feature appearance from high-quality or even space-based observations can provide a quality metric which objectively estimates the distortion of solar features. (2) The extension to a metric which can operate between observations from \textbf{different instruments} requires the further extension of the data set, in order to compensate for the instrumental differences, while it still needs to give an objective quality estimation based on the highest image quality attainable. In a first test, the pre-trained model was applied to overlapping data series from KSO and Uccle Solar Equatorial Table (USET), and demonstrated the capability to filter low-quality images between different observation sites. To account for small quality variations between different sites, higher resolutions should be included to allow for the dynamic selection of the highest image quality. (3) A different approach is the \textbf{detection of solar transient events}. Our method shows already some sensitivity for flares. By removing all flaring samples from the training set, a detection module could be obtained. We further note that this concept is not restricted to image data, but could also be applied to 1D or 3D data.

\section{Conclusions}
We presented a method for the stable classification and quantification of image quality in ground-based solar full-disk images. From a set of regular observations, we derived an objective no-reference image quality metric which accounts for seeing, atmospheric effects and instrumental errors in H$\alpha$ filtergrams. Our method achieved an almost perfect detection of anomalies (99.5\%) for an unfiltered time series of several observing days (covering in total 10,050 images), and a high performance on an independent test set for the years 2012 - 2019 (3,300 images) covering a broad variety of atmospheric effects and solar activity. Our top performing neural network achieved an accuracy of 98.5\% and a TSS of 0.98 in separating high-quality observations from observations with degrading effects (low-quality), as compared to visual inspection. The proposed image quality metric shows a good agreement with the human perception and provides a smooth transition between high-quality and low-quality observations. Once the model is trained, it can be operated without any further reference image. The processing time is short (about 17 ms for inference) as compared to typical observing cadences and can easily be parallelized, allowing for efficient real time application.

Our method is based on two important concepts that make our approach superior to existing methods. (1) We made use of the true image distribution and quantify deviations from it. (2) We employ an image quality metric which estimates feature similarity rather than pixel-based variations. Since the model can be trained solely with the use of regular observations, it is suitable for many applications in observational astrophysics. With the availability of a sufficient amount of low-quality observations, the model performance can be further increased by using the proposed classifier architecture. In addition to the quality score, our method provides an identification of the image regions affected by reduced quality. The presented method can be adapted to other instruments, wavelength channels or observing targets. Furthermore, our method offers potential for the application to high-resolution solar physics data, homogenization of observation series from telescope networks, time series and event detection.

\begin{acknowledgements}
This research has received financial support from the European Union’s Horizon 2020 research and innovation program under grant agreement No. 824135 (SOLARNET).
The computational results presented have been achieved using the Vienna Scientific Cluster (VSC) and the Skoltech HPC cluster ARKUDA.
This research has made use of SunPy v1.1.4 \citep{sunpy_software2020}, an open-source and free community-developed solar data analysis Python package \citep{sunpy_community2020}.

\end{acknowledgements}

\bibliographystyle{aa}
\bibliography{references}

\begin{thebibliography}{40}
\expandafter\ifx\csname natexlab\endcsname\relax\def\natexlab#1{#1}\fi

\bibitem[{Agustsson {et~al.}(2019)Agustsson, Tschannen, Mentzer, Timofte, \&
  Gool}]{agustsson2019generative}
Agustsson, E., Tschannen, M., Mentzer, F., Timofte, R., \& Gool, L.~V. 2019, in
  Proceedings of the IEEE International Conference on Computer Vision, 221--231

\bibitem[{Barnes {et~al.}(2016)Barnes, Leka, Schrijver, Colak, Qahwaji,
  Ashamari, Yuan, Zhang, McAteer, Bloomfield, {et~al.}}]{barnes2016comparison}
Barnes, G., Leka, K., Schrijver, C., {et~al.} 2016, The Astrophysical Journal,
  829, 89

\bibitem[{Barnes {et~al.}(2020)Barnes, Bobra, Christe, Freij, Hayes, Ireland,
  Mumford, Perez-Suarez, Ryan, Shih, {et~al.}}]{sunpy_community2020}
Barnes, W.~T., Bobra, M.~G., Christe, S.~D., {et~al.} 2020, The Astrophysical
  Journal, 890, 68

\bibitem[{Blau \& Michaeli(2018)}]{blau2018perceptionDistortion}
Blau, Y. \& Michaeli, T. 2018, in Proceedings of the IEEE Conference on
  Computer Vision and Pattern Recognition, 6228--6237

\bibitem[{Chambolle(2004)}]{chambolle2004algorithm}
Chambolle, A. 2004, Journal of Mathematical imaging and vision, 20, 89

\bibitem[{Chollet(2017)}]{chollet2017xception}
Chollet, F. 2017, in Proceedings of the IEEE conference on computer vision and
  pattern recognition, 1251--1258

\bibitem[{Deng {et~al.}(2015)Deng, Zhang, Wang, Ji, Wang, Liu, Xiang, Jin, \&
  Cao}]{deng2015objective}
Deng, H., Zhang, D., Wang, T., {et~al.} 2015, Solar Physics, 290, 1479

\bibitem[{Galvez {et~al.}(2019)Galvez, Fouhey, Jin, Szenicer,
  Mu{\~n}oz-Jaramillo, Cheung, Wright, Bobra, Liu, Mason,
  {et~al.}}]{galvez2019machine}
Galvez, R., Fouhey, D.~F., Jin, M., {et~al.} 2019, The Astrophysical Journal
  Supplement Series, 242, 7

\bibitem[{Goodfellow {et~al.}(2016)Goodfellow, Bengio, \&
  Courville}]{goodfellow2016deep}
Goodfellow, I., Bengio, Y., \& Courville, A. 2016, Deep learning (MIT press)

\bibitem[{Goodfellow {et~al.}(2014{\natexlab{a}})Goodfellow, Pouget-Abadie,
  Mirza, Xu, Warde-Farley, Ozair, Courville, \& Bengio}]{goodfellow2014gan}
Goodfellow, I., Pouget-Abadie, J., Mirza, M., {et~al.} 2014{\natexlab{a}}, in
  Advances in neural information processing systems, 2672--2680

\bibitem[{Goodfellow {et~al.}(2014{\natexlab{b}})Goodfellow, Shlens, \&
  Szegedy}]{goodfellow2014explaining}
Goodfellow, I.~J., Shlens, J., \& Szegedy, C. 2014{\natexlab{b}}, arXiv
  preprint arXiv:1412.6572

\bibitem[{Gosain {et~al.}(2018)Gosain, Roth, Hill, Pevtsov, Pillet, \&
  Thompson}]{gosain2018design}
Gosain, S., Roth, M., Hill, F., {et~al.} 2018, in Ground-based and Airborne
  Instrumentation for Astronomy VII, Vol. 10702, International Society for
  Optics and Photonics, 107024H

\bibitem[{Grundahl {et~al.}(2006)Grundahl, Kjeldsen, Frandsen, Andersen,
  Bedding, Arentoft, \& Christensen-Dalsgaard}]{grundahl2006song}
Grundahl, F., Kjeldsen, H., Frandsen, S., {et~al.} 2006, A global network of
  small telescopes for asteroseismology and planet searches. Memorie della
  Societa Astronomica Italiana, 77, 458

\bibitem[{Harvey {et~al.}(1996)Harvey, Hill, Hubbard, Kennedy, Leibacher,
  Pintar, Gilman, Noyes, Toomre, Ulrich, {et~al.}}]{harvey1996global}
Harvey, J., Hill, F., Hubbard, R., {et~al.} 1996, Science, 272, 1284

\bibitem[{He {et~al.}(2016)He, Zhang, Ren, \& Sun}]{he2016deep}
He, K., Zhang, X., Ren, S., \& Sun, J. 2016, in Proceedings of the IEEE
  conference on computer vision and pattern recognition, 770--778

\bibitem[{Huang {et~al.}(2019)Huang, Jia, Cai, \& Cai}]{huang2019perception}
Huang, Y., Jia, P., Cai, D., \& Cai, B. 2019, Solar Physics, 294, 133

\bibitem[{Isola {et~al.}(2017)Isola, Zhu, Zhou, \& Efros}]{isola2017image}
Isola, P., Zhu, J.-Y., Zhou, T., \& Efros, A.~A. 2017, in Proceedings of the
  IEEE conference on computer vision and pattern recognition, 1125--1134

\bibitem[{Johnson {et~al.}(2016)Johnson, Alahi, \&
  Fei-Fei}]{johnson2016perceptual}
Johnson, J., Alahi, A., \& Fei-Fei, L. 2016, in European conference on computer
  vision, Springer, 694--711

\bibitem[{Karras {et~al.}(2017)Karras, Aila, Laine, \&
  Lehtinen}]{karras2017progressive}
Karras, T., Aila, T., Laine, S., \& Lehtinen, J. 2017, arXiv preprint
  arXiv:1710.10196

\bibitem[{Kingma \& Ba(2014)}]{kingma2014adam}
Kingma, D.~P. \& Ba, J. 2014, arXiv preprint arXiv:1412.6980

\bibitem[{LeCun {et~al.}(2015)LeCun, Bengio, \& Hinton}]{lecun2015deep}
LeCun, Y., Bengio, Y., \& Hinton, G. 2015, nature, 521, 436

\bibitem[{{L{\"o}fdahl} {et~al.}(2007){L{\"o}fdahl}, {van Noort}, \&
  {Denker}}]{lofdahl2007restoration}
{L{\"o}fdahl}, M.~G., {van Noort}, M.~J., \& {Denker}, C. 2007, in Modern solar
  facilities - advanced solar science, ed. F.~{Kneer}, K.~G. {Puschmann}, \&
  A.~D. {Wittmann}, 119

\bibitem[{Mao {et~al.}(2017)Mao, Li, Xie, Lau, Wang, \&
  Paul~Smolley}]{mao2017least}
Mao, X., Li, Q., Xie, H., {et~al.} 2017, in Proceedings of the IEEE
  International Conference on Computer Vision, 2794--2802

\bibitem[{Mentzer {et~al.}(2018)Mentzer, Agustsson, Tschannen, Timofte, \&
  Van~Gool}]{mentzer2018conditional}
Mentzer, F., Agustsson, E., Tschannen, M., Timofte, R., \& Van~Gool, L. 2018,
  in Proceedings of the IEEE Conference on Computer Vision and Pattern
  Recognition, 4394--4402

\bibitem[{Mittal {et~al.}(2012)Mittal, Moorthy, \& Bovik}]{mittal2012brisque}
Mittal, A., Moorthy, A.~K., \& Bovik, A.~C. 2012, IEEE Transactions on image
  processing, 21, 4695

\bibitem[{Mumford {et~al.}(2020)Mumford, Christe, Freij, Mayer, Hughitt, Ryan,
  Liedtke, Shih, Pérez-Suárez, Chakraborty, I, Inglis, Pattnaik, Sipőcz,
  Sharma, Leonard, Hewett, Alex-Ian-Hamilton, Stansby, Panda, Earnshaw,
  Choudhary, Kumar, Hayes, Chanda, Haque, Konge, mdmueller, Kirk, \&
  haathi}]{sunpy_software2020}
Mumford, S.~J., Christe, S., Freij, N., {et~al.} 2020, SunPy

\bibitem[{{Otruba} \& {P{\"o}tzi}(2003)}]{otruba2003halphaKSO}
{Otruba}, W. \& {P{\"o}tzi}, W. 2003, Hvar Observatory Bulletin, 27, 189

\bibitem[{Papernot {et~al.}(2017)Papernot, McDaniel, Goodfellow, Jha, Celik, \&
  Swami}]{papernot2017practical}
Papernot, N., McDaniel, P., Goodfellow, I., {et~al.} 2017, in Proceedings of
  the 2017 ACM on Asia conference on computer and communications security,
  506--519

\bibitem[{{Pesnell} {et~al.}(2012){Pesnell}, {Thompson}, \&
  {Chamberlin}}]{pesnell2012sdo}
{Pesnell}, W.~D., {Thompson}, B.~J., \& {Chamberlin}, P.~C. 2012, \solphys,
  275, 3

\bibitem[{Popowicz {et~al.}(2017)Popowicz, Radlak, Bernacki, \&
  Orlov}]{popowicz2017review}
Popowicz, A., Radlak, K., Bernacki, K., \& Orlov, V. 2017, Solar Physics, 292,
  187

\bibitem[{P{\"o}tzi {et~al.}(2018)P{\"o}tzi, Veronig, \&
  Temmer}]{potzi2018event}
P{\"o}tzi, W., Veronig, A., \& Temmer, M. 2018, Solar Physics, 293, 94

\bibitem[{P{\"o}tzi {et~al.}(2015)P{\"o}tzi, Veronig, Riegler, Amerstorfer,
  Pock, Temmer, Polanec, \& Baumgartner}]{potzi2015real}
P{\"o}tzi, W., Veronig, A.~M., Riegler, G., {et~al.} 2015, Solar physics, 290,
  951

\bibitem[{{Rimmele} \& {Marino}(2011)}]{rimelle2011solaradaptive}
{Rimmele}, T.~R. \& {Marino}, J. 2011, Living Reviews in Solar Physics, 8, 2

\bibitem[{Simonyan \& Zisserman(2014)}]{simonyan2014very}
Simonyan, K. \& Zisserman, A. 2014, arXiv preprint arXiv:1409.1556

\bibitem[{{Veronig} \& {P{\"o}tzi}(2016)}]{veronig2016spaceweather}
{Veronig}, A.~M. \& {P{\"o}tzi}, W. 2016, in Astronomical Society of the
  Pacific Conference Series, Vol. 504, Coimbra Solar Physics Meeting:
  Ground-based Solar Observations in the Space Instrumentation Era, ed.
  I.~{Dorotovic}, C.~E. {Fischer}, \& M.~{Temmer}, 247

\bibitem[{Wang {et~al.}(2018)Wang, Liu, Zhu, Tao, Kautz, \&
  Catanzaro}]{wang2018pix2pixhd}
Wang, T.-C., Liu, M.-Y., Zhu, J.-Y., {et~al.} 2018, in Proceedings of the IEEE
  conference on computer vision and pattern recognition, 8798--8807

\bibitem[{Wang {et~al.}(2004)Wang, Bovik, Sheikh, \&
  Simoncelli}]{wang2004image}
Wang, Z., Bovik, A.~C., Sheikh, H.~R., \& Simoncelli, E.~P. 2004, IEEE
  transactions on image processing, 13, 600

\bibitem[{{W{\"o}ger} {et~al.}(2008){W{\"o}ger}, {von der L{\"u}he}, \&
  {Reardon}}]{woger2008speckle}
{W{\"o}ger}, F., {von der L{\"u}he}, O., \& {Reardon}, K. 2008, \aap, 488, 375

\bibitem[{Ye {et~al.}(2012)Ye, Kumar, Kang, \& Doermann}]{ye2012cornia}
Ye, P., Kumar, J., Kang, L., \& Doermann, D. 2012, in 2012 IEEE conference on
  computer vision and pattern recognition, IEEE, 1098--1105

\bibitem[{Zhou {et~al.}(2016)Zhou, Khosla, Lapedriza, Oliva, \&
  Torralba}]{zhou2016cnnfeatures}
Zhou, B., Khosla, A., Lapedriza, A., Oliva, A., \& Torralba, A. 2016, in
  Proceedings of the IEEE conference on computer vision and pattern
  recognition, 2921--2929

\end{thebibliography}

%
%
\begin{appendix}

\section{Model Architecture}
\label{section:model_architecture}
Our primary model is composed of an encoder, quanitzer and decoder. We use the notation from \cite{johnson2016perceptual} and \cite{wang2018pix2pixhd}, where c\textbf{k}s\textbf{j}-\textbf{n} denotes a \textbf{k}x\textbf{k} Convolution Layer with stride \textbf{j}, Instance-Normalization and ReLU activation with \textbf{n} filters. d\textbf{k} refers to a 3x3 Convolution with stride 2, Instance-Normalization and ReLU activation with \textbf{k} filters, whereas u\textbf{k} denotes the same configuration with Transposed Convolution instead of Convolution Layers. R\textbf{n} refers to a residual block as proposed in the ResNet50 architecture, with \textbf{n} filters \citep{he2016deep}. Throughout our network we use reflection padding, in order to reduce boundary artifacts \citep{wang2018pix2pixhd}. The last convolution layer of the encoder uses the activation of the quantizer instead of a ReLU activation (see Sect. \ref{section:quantizer}).

The discriminator uses 4 consecutive stride 2 convolutions with instance normalization and a leaky-relu activation with a slope of 0.2 which we denote with C\textbf{n}, where \textbf{n} refers to the number of filters. No normalization is applied to the first convolutional layer \citep{wang2018pix2pixhd}. Per convolution the amount of filters is increased by a factor of 2, while the spatial dimension is reduced by a factor of 4. The output is produced by a final convolutional layer where we omit the normalization and use a tanh activation.

For the classifier network, we apply the same architecture as for the discriminator. We use 3 discriminators and 3 classifiers, where we use the full resolution as provided by the generator for the first discriminator/classifier and consecutively reduced resolutions for the second and third. For this task, we apply average pooling to the input images.

\textbf{Generator} (with number of quantization channels y):\\
Encoder: c7s1-64,d128,d256,d512,c3s1-y,Q\\
Decoder: 9x R512,u256,u128,u64,c7s1-1\\

\textbf{Discriminator/Classifier} (3x with Average-Pooling of 1, 2 and 4):
C64,C128,C256,C512

\section{Quantizer}
\label{section:quantizer}
For the discrete representation $\hat{\omega}$ we use the quantizer $Q$ as proposed in \cite{agustsson2019generative} which uses a hard non-differentiable quantization in the forward pass and a differentiable approximation in the backward pass of model training. This is implemented with a gradient stop:
\begin{equation}
    \hat{\omega} =  $tf.stop\_gradient$(\hat{z} - \Tilde{z}) + \Tilde{z}.
\end{equation}
 Here the hard quantization $\hat{z}$ is computed by rounding the output of the encoder $z$ to integers and the soft quantization $\Tilde{z}$ is obtained by applying the softmax function to the absolute difference between the encoder output $z$ and the discrete centers $c$ \citep{mentzer2018conditional}:

\begin{equation}
     \hat{\omega}_i = \sum_{j=1}^L \frac{\exp(-\Vert z_i - c_j  \Vert_1)}{\sum_{l=1}^L \exp(-\Vert z_i - c_l \Vert_1)}  c_j.
\end{equation}
The encoder output $z$ is calculated from the last features in the encoder:
\begin{equation}
    z_i = \sigma(x_i)*L,
\end{equation}
where $x$ refers to the output of the last convolutional layer in the encoder, $L$ to the number of centers and sigma to the sigmoid activation function.

\section{Samples}
\begin{figure*}[b]%
\includegraphics[width=\linewidth]{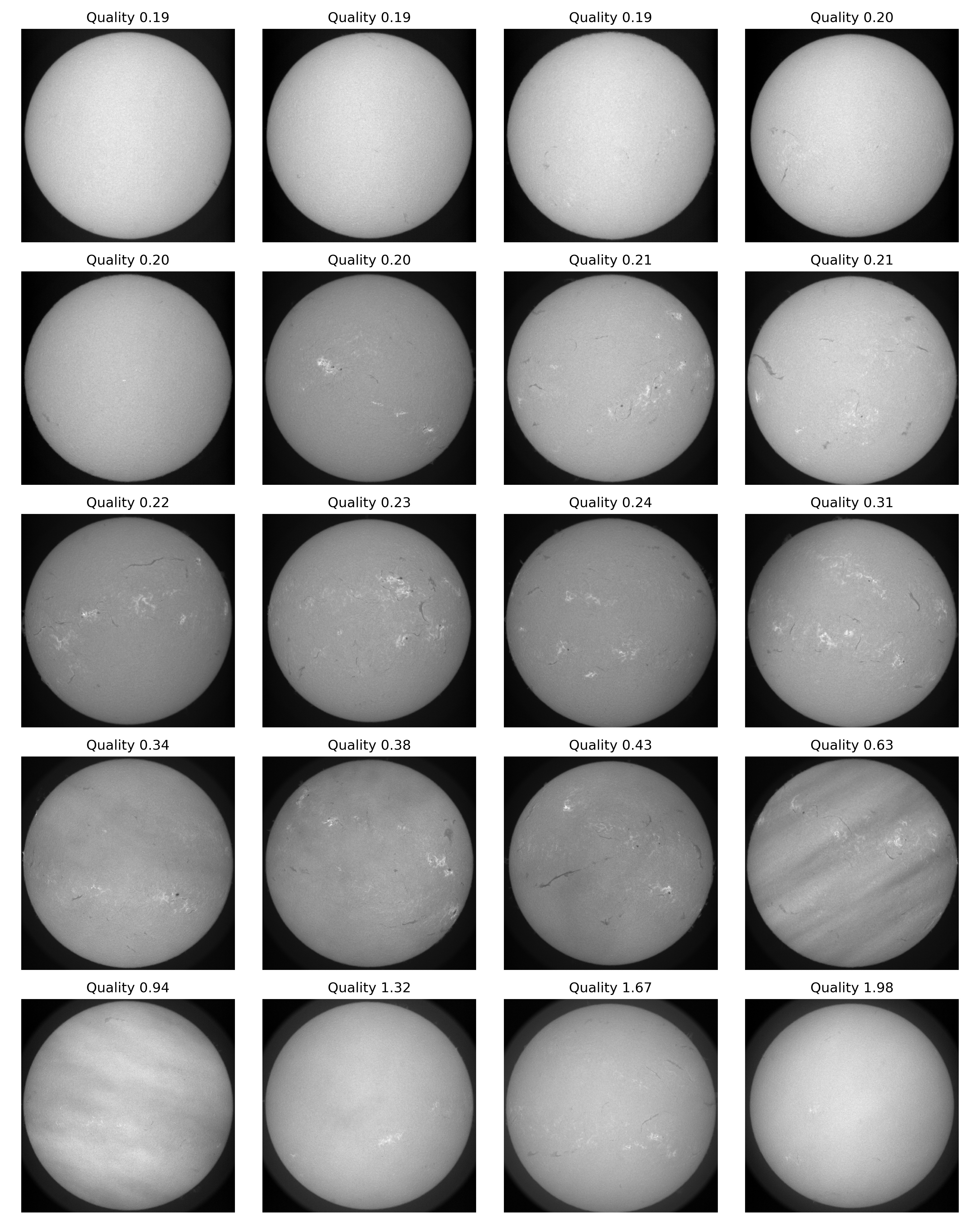}
\caption{Examples from the test set with adjustment of the off-limb region (see Sect. \ref{section:metric}). The images are sampled across the full test set and sorted with respect to the estimated image quality value.}
\label{figure:samples_grid}
\end{figure*}

In Fig. \ref{figure:samples_grid} we show examples of different quality degradation. The image quality, as estimated by our CLASS-q8-CONTR model, is indicated on top of the images.

\end{appendix}
\end{document}